\documentclass{cup-hpl}

\usepackage{lineno}

\begin{document}

\newtheorem{theorem}{Theorem}

\shorttitle{HRR Exploration of Biermann Battery Effect in LPPs over Large Spatial Regions}                                   
\shortauthor{J.~J. Pilgram et al.}

\title{High Repetition Rate Exploration of the Biermann Battery Effect in Laser Produced Plasmas Over Large Spatial Regions}

\author[1]{J.J. Pilgram\corresp{Address of corresp.
                       \email{jpilgram@ucla.edu}}}
\author[2]{M.~B.~P. Adams}
\author[1]{C.~G. Constantin}
\author[3]{P.~V. Heuer}
\author[1]{S. Ghazaryan}
\author[1]{M. Kaloyan}
\author[1]{R.~S. Dorst}
\author[4]{D.~B. Schaeffer}
\author[2,3]{P. Tzeferacos}
\author[1]{C. Niemann}

\address[1]{Department of Physics and Astronomy, University of California, Los Angeles, Los Angeles, CA 90095, USA}
\address[2]{Department of Physics and Astronomy, University of Rochester, Rochester, NY 14627, USA}
\address[3]{Laboratory for Laser Energetics, University of Rochester, Rochester, NY, 14623, USA}
\address[4]{Department of Astrophysical Sciences, Princeton University, Princeton, NJ 08540, USA}

\begin{abstract}
In this paper we present a high-repetition-rate experimental platform for examining the spatial structure and evolution of Biermann generated magnetic fields in laser-produced plasmas. We have extended the work of prior experiments, which spanned over millimeter scales, by spatially measuring magnetic fields in multiple planes on centimeter scales over thousands of laser shots. Measurements with magnetic flux probes show azimuthally symmetric magnetic fields that range from 60~G at 0.7~cm from the target to 7~G at 4.2~cm from the target. The expansion rate of the magnetic fields and evolution of current density structures are also mapped and examined. Electron temperature and density of the laser-produced plasma are measured with optical Thomson scattering and used to directly calculate a magnetic Reynolds number of $1.4\times 10^4$, confirming that magnetic advection is dominant $\ge 1.5$ cm from the target surface. The results are compared to FLASH simulations, which show qualitative agreement with the data.
\end{abstract}


\maketitle

\section{\label{sec:level1}Introduction}
Magnetic fields are prevalent throughout the universe. The  generation of cosmic magnetism represents an important problem in modern astrophysics. As described by Kulsrud and Zweibel \cite{Kulsrud2008On}, the cosmological evolution of the universe cannot be fully understood without solid knowledge of the origin of magnetic fields, structures, and evolution. These fields are hard to detect as they are very weak (in the micro-Gauss range) and far away\cite{Kulsrud2008On,Rand1989The, Beck2007Magnetic, Han2006Pulsar}. Diagnostic techniques such as Faraday-rotation and Zeeman splitting are difficult to implement in such contexts\cite{Zweibel1997Magnetic,Heiles1998Zeeman,Kulsrud2008On}. However, laboratory astrophysics experiments, which reproduce astrophysical plasmas scaled by dimensionless parameters, can supplement observational measurements by addressing these limitations\cite{Remington2000A,Ryutov2001MHD}. Thus, the marriage of astrophysical observation and theory, laboratory experimentation using laser-produced plasmas, and computational modeling of such scenarios, aid in the pursuit of answering questions on cosmic magnetic field generation. 

One predominant theory for primordial cosmic magnetic seed fields is the Biermann battery mechanism\cite{Kulsrud2008On,Zweibel2013,Naoz2013Generation,Gregori2012Generation}, which is a thermo-electric process that spontaneously generates magnetic fields in plasmas via non-parallel temperature and density gradients. This effect was first described by Ludwig Biermann in 1949 \cite{Biermann1949} and has been studied in a variety of plasma experiments \cite{Stamper1971Spontaneous,Pisarczyk2015Space-time,Gopal2008Temporally,Mckee1974Self-generated,Stamper1991Review, Gao2015Precision} due to its importance not only for cosmic fields but also for its influential role in many laboratory plasma phenomena, such as magnetic reconnection\cite{Nilson2006Magnetic}, laboratory shocks\cite{Gregori2012Magnetic, Gregori2012Generation}, and components of inertial confinement fusion schemes\cite{Li2008Monoenergetic,Walsh2017Self-Generated}. 

Laser-produced plasma (LPP) platforms are commonly used to study these spontaneously generated magnetic fields\cite{Gao2015Precision,Campbell2020Magnetic}. Not only do LPPs naturally produce the density and temperature gradients needed to precipitate the Biermann battery, but the magnetic fields generated in laser experiments by the Biermann battery mechanism play a significant role in the energy transport in plasmas, affecting particle dynamics\cite{Braginskii1965Transport}, leading to hot spots\cite{Craxton1975Hot}, fast electrons and ions\cite{Craxton1978JxB}, and producing large magnetic pressures (i.e., high plasma-$\beta$))\cite{Raven1978Megagauss,Haines1986Magnetic}.

The amplitudes of laboratory generated Biermann fields range from weak (a few micro-Gauss) to very strong (mega-Gauss)\cite{Raven1978Megagauss,Haines1986Magnetic} and have shown a direct dependence on the laser irradiation intensity in the range of 10$^{12}$-10$^{14}$~W/cm$^2$. McLean et al. \cite{McLean1984Observation} detected up to 300~kG fields; Stamper et al. demonstrated MG fields at 10$^{15}$~W/cm$^2$, as did Pisarczyk et al. \cite{Pisarczyk2015Space-time} and Gopal et al. \cite{Gopal2008Temporally} in plasmas generated by lasers with intensities above 10$^{16}$~W/cm$^2$. In all cases, the peak amplitudes were found very close to the target surfaces, at distances of less than 1~cm. A two dimensional map of the Biermann battery effect was generated by McKee et al. \cite{Mckee1974Self-generated} extending 2~cm from the target surface. These measurements also confirmed the findings of Bird et al. \cite{Bird1973Pressure} that stronger fields may be generated in the presence of a background gas.

Although this large body of work establishes a foundation for understanding the process of magnetic field formation, there is a lack of data on how Biermann fields behave at larger spatial scales, including the range over which magnetic advection or diffusion dominate. In this paper we present a new high-repetition-rate (HRR) experimental platform for studying the generation and evolution of Biermann magnetic fields in laser-produced plasmas over large (tens of cm) spatial scales. By combining a HRR laser driver and motorized magnetic flux probe, we obtain over thousands of shots high-spatial-resolution, three-dimensional maps of the evolution of Biermann generated magnetic fields and current density structures.  Additional HRR measurements with an optical Thomson scattering probe beam allow us to measure plasma density and temperature. From this data we directly calculate the magnetic Reynolds number and show that magnetic advection dominates at distances $\ge 1.5$~cm from the target. Finally, we compare our results to preliminary FLASH simulations modeled after the experiments, which show qualitative agreement with the data.\footnote{For more information on the FLASH code, visit: \url{http://flash.rochester.edu}}.

\section{\label{theory}Theoretical Background}
The evolution of the magnetic fields in a non-ideal resistive magnetohydrodynamics framework is described by an induction equation of the form, 
\begin{align}
\nonumber \frac{\partial \vec{B}}{\partial t} = \vec{\nabla}\times& (\vec{v_e}\times\vec{B}) + \frac{\eta c^2}{4\pi}\nabla^2\vec{B} \\
        &-\frac{1}{en_e}\vec{\nabla}\times (\vec{J} \times \vec{B}) + \frac{c}{en_e}\vec{\nabla}T_e\times\vec{\nabla}n_e.
    \label{Ohm}
\end{align}

\noindent where $\vec{B}$ is the magnetic field, $\vec{v_e}$ is the electron velocity, $\eta$ is the plasma resistivity, $c$ is the speed of light, $\vec{J}$ is the current density, $T_e$ is the electron temperature $n_e$ is the electron number density, and $e$ is the electron charge. The first term on the right hand side of Eq. \ref{Ohm} describes the convection of the magnetic fields in the plasma, the second term denotes the diffusion of the magnetic field with respect to the plasma, the third term is the Hall term which describes the redistribution of the magnetic fields due to Hall forces, and the fourth term is the magnetic source term, also known as the Biermann battery term. 

In LPPs, the primary temperature gradient is perpendicular to the axis of the plasma plume and the primary density gradient is normal to the target surface (with higher density closer to the target). The electrons collectively move parallel with the pressure gradient at higher velocities than the heavier ions. This action generates an electromotive force (EMF). By Faraday's law, this EMF in turn creates a magnetic flux, and thus a magnetic field is spontaneously created in an azimuthal direction around the plasma blow off axis. The term in the induction equation that is of interest in this study is the source term,

\begin{equation}
    \vec{B}_{source} = -\frac{c}{en_e}\left( \frac{\partial T_{e}}{\partial r}\frac{\partial n_{e}}{\partial z} - \frac{\partial T_{e}}{\partial z}\frac{\partial n_{e}}{\partial r}\right) \hat{\theta},
    \label{B_source}
\end{equation}

\noindent where $\hat{\theta}$ is the azimuthal unit vector. Note that in a cylindrical framework, $r=\sqrt{x^{2}+y^{2}}$ is the radial coordinate, and $z$ is the axial coordinate. Thus derivatives in the radial and axial directions for the electron temperature and density are crucial to our understanding of Biermann fields generated in the LPP context. How we measure such azimuthal fields is discussed in the following section.

To determine if diffusion of the magnetic field dominates over advection in our plasma, we estimate the magnetic Reynolds number. The magnetic Reynolds number is found by non-dimensionalizing Eq.~\ref{Ohm}. This number is a dimensionless ratio between the magnetic advection and diffusion within a plasma and is defined as,

\begin{equation}
R_m = UL/\eta
\label{Rm}
\end{equation}

\noindent where $U$ is the velocity scale of plasma flow, $L$ is the length scale of the plasma flow, and $\eta$ is the magnetic diffusivity. $R_m\gg1$ implies that advection dominates, where as $R_m\ll1$ implies that diffusion dominates.

\section{\label{ExpDesign} Experimental Design}
\begin{figure}
\centering
  \includegraphics[width=1.0\linewidth]{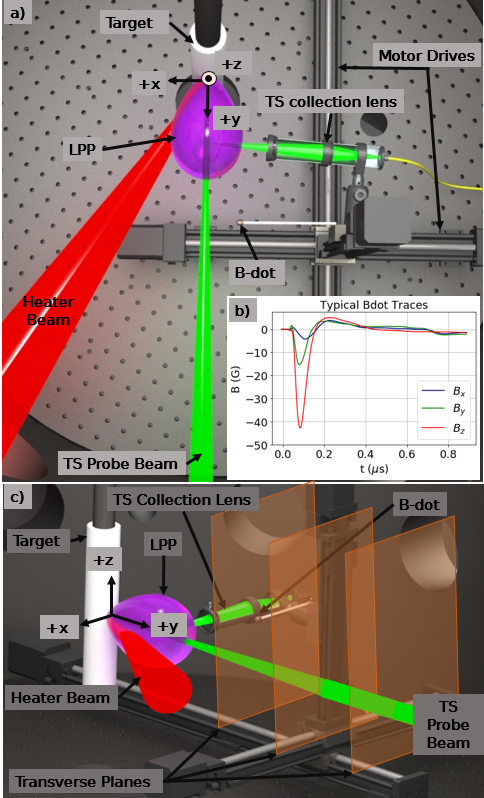}
\caption{A rendering of the experimental setup. \textbf{a)} Top view. The origin of the coordinate system is the laser spot on target, with the corresponding axis directions as depicted. \textbf{b)} Typical B-dot probe traces for all three axes of the probe. \textbf{c)} Side view. The translucent orange rectangles represent the planes in which magnetic field data was collected.}
\label{ExpSetUp}
\end{figure}

The experimental setup is illustrated in Figure \ref{ExpSetUp}. The experiment took place inside a one meter diameter, stainless steel, cylindrical vacuum chamber. LPPs were created by irradiating a 25~mm diameter cylindrical high-density polyethylene (C$_2$H$_4$) target with a pulsed high-energy heater laser at 34~degree incidence angle with respect to the target surface normal. The heater laser was a Nd:glass system at 1053~nm wavelength, 15~ns pulse length, with a repetition rate of up to 6~Hz, and a maximum output energy of 20~J. The eight-pass amplification scheme involves a phase-conjugation wavefront reversal technique to output a near diffraction-limited spot~\cite{Dane1995Peening}. In these experiments, the heater laser energy was 10~J and the repetition rate was 1~Hz. The laser was controlled and monitored through a custom-built LabView application that allows automatic synchronization of the laser with the diagnostics and the target motion systems at 1~Hz. The laser was focused down to a 250~$\mu$m diameter spot onto the target by an f/25 lens, yielding a nominal intensity of $I \approx 1.3 \times 10^{12} W/cm^2$.

Measurements of the magnetic field flux were collected using a three-axis magnetic flux (``B-dot'') probe, the design and construction of which is described in detail elsewhere~\cite{Everson2009Design}. The B-dot probe consisted of three sets of thin wire coils wound around three perpendicular axes each with a diameter of 3~mm.  
When a magnetic flux passes through the wire coils in the probe, a current is induced. This current is then passed through a differential amplifier and recorded using a digitizer (250~MSamples/s, 125~MHz bandwidth, 12~bit), which recorded for 3.5~$ \mu s$ after the laser shot.
The recorded voltage signals are then integrated to yield magnetic field measurements. The integration method used is described in section \ref{Results}.

The B-dot probe was mounted on an automated 3-axis stepper motor-driven stage inside the chamber.  
An extensive volumetric scan comprising thousands of shots was created by moving the probe in a pre-defined 3D pattern at a 1~Hz rate with spatial steps of $2 \pm 0.05$~mm.
The target was rotated and translated vertically in a helical pattern in order to ensure a fresh surface for each shot. Five shots were taken at each spatial position for data averaging in order to account for shot-to-shot laser intensity fluctuations (approximately $5\%$ per shot\cite{Schaeffer2018A}) and small shot-to-shot fluctuations in the background noise detected by the B-dot probe. The full spatial scanning capability of the motor drive set up is $[-7,-63]$~mm in the $x$ direction, $[7,139]$~mm in the $y$ direction, and $[-85,85]$~mm in the $z$ direction, with the laser spot representing the origin of the coordinate system. The closest distance to the target for the B-dot probe was 7~mm, limited by proximity to the laser path. 

Due to the direction of the generated fields, we focus on several planes which are perpendicular to the plasma blow-off axis such that we can observe the azimuthal structure of the generated fields. By combining many of these perpendicular planes, we are able to create a three-dimensional picture of the measured Biermann fields. 
In this experiment, data was collected at the same x and z points in planes at various distances from the target surface along the plasma blow-off axis (y-axis), as illustrated in Figure \ref{ExpSetUp}. Each plane was separated from surrounding planes by a spatial distance of 5~mm.

The plasma temperature and density was measured using optical Thomson scattering~\cite{Froula2012Plasma}. For these measurements, the plasma was probed by a separate 532~nm wavelength probe laser with a 50~mJ output energy and 4~ns pulse width, at 1~Hz repetition rate. The probe laser enters the chamber along the plasma blow-off axis ($y$-axis) and terminates on the target. The small secondary laser plasma created by this probe laser reaches the B-dot probe and Thomson scattering volume hundreds of ns after the time of data collection and thus does not affect the results. A detailed description of the Thomson scattering setup and measurements have been published separately\cite{Kaloyan2021Raster}. 

\section{\label{Results} Results and Analysis}

Magnetic flux measurements were taken in multiple planes at varying distances from the target surface. The magnetic flux was calculated from the measured voltage traces using the following integration method which is described by Everson et al.~\cite{Everson2009Design}:
\begin{equation}
\vec{B} =  \frac{A}{an_bg} \left[ \int{\vec{V}_{measured}(t)dt + \tau_s\vec{V}_{measured}(t)+ \vec{V_o}} \right] + \vec{B_o}  
\end{equation}

\noindent where $g = 10$ is the amplifier gain, $A$ is the attenuation factor which varies depending on the signal strength, $n_b = 10$ is the number of turns in the coils, $a$ is the area of the coils determined by the calibration ($a_x \approx 11.3$ mm$^2$, $a_y \approx 9.1$ mm$^2$, $a_z \approx 13.2$ mm$^2$), $\vec{B_o} = 0$ is the initial background field, $V_0$ is the background noise which was previously subtracted, and $\tau_s$ is the time constant associated with the RL circuit formed by the coils ($\tau \approx 30\ ns$). 
The error of the resultant magnetic fields was calculated by taking the standard deviation of the five shots taken at each spatial position.
The data was then averaged over five shots for a better signal-to-noise ratio and to account for the shot-to-shot laser energy fluctuations.

\begin{figure*}
    \centering
    \includegraphics[width = 1.0\textwidth]{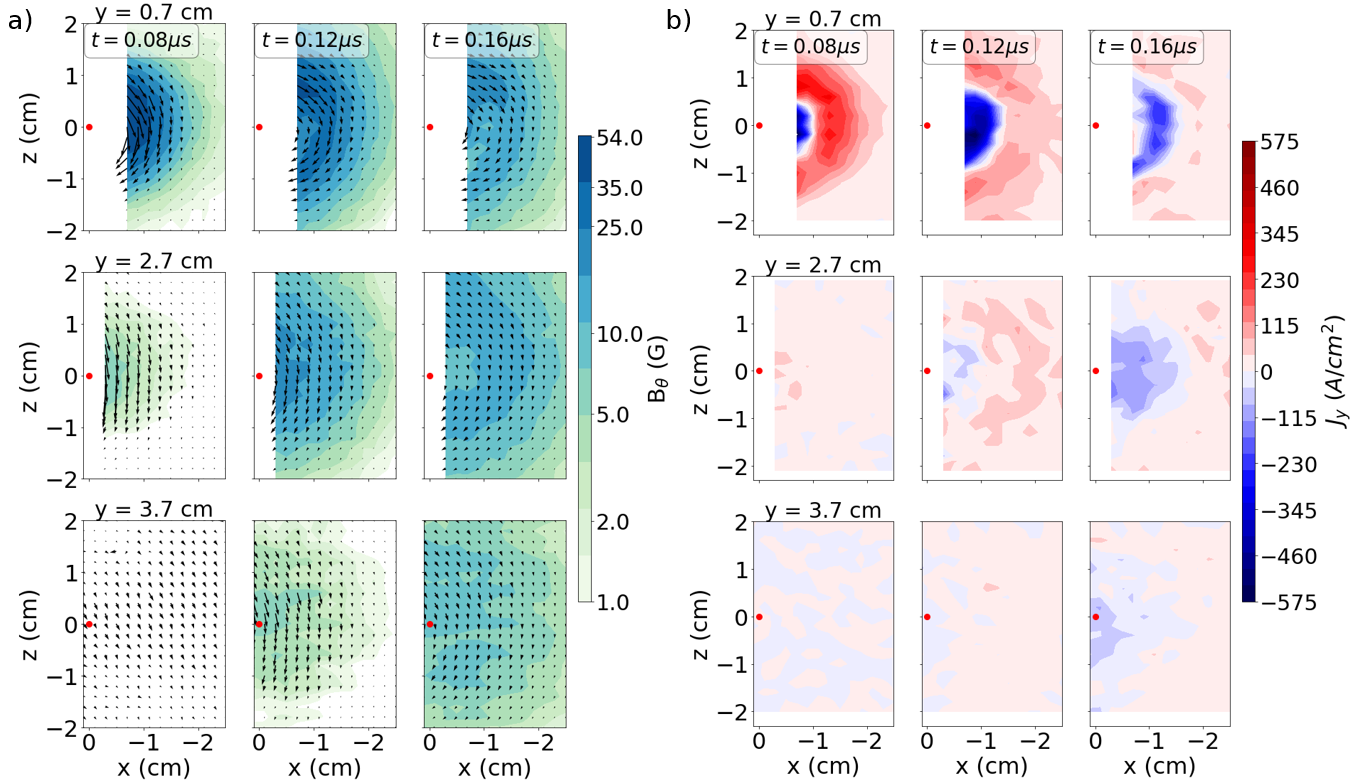}
    \caption{\textbf{a)} Contour plots of azimuthal magnetic field structure in several transverse planes at three representative times. Magnetic field vectors are denoted by the black arrows. \textbf{b)} The calculated current density along the plasma blow-off axis in several transverse planes at three representative times. The red dot represents the laser spot and white spaces are positions which the probe could not reach due to mechanical constraints.}
    \label{Planes}
\end{figure*}

The position of the B-dot probe was scanned in many transverse planes at distances between 7 and 42 mm from the target surface, each separated by a distance of 5~mm. These planes are pieced together to allow for analysis of the magnetic field structure in three-dimensions. A representative portion of the magnetic field data is presented in Figure \ref{Planes}a. This figure depicts three data planes showing the detected azimuthal magnetic fields, $B_{\theta}$, at three distances from the target surface. 
The contour values represent the magnitude of $B_{\theta}$ and black vectors superimposed on the plots indicate the magnetic field orientation, as measured by the B-dot probe. In these planes, the LPP propagates out of the page. The temperature gradient points towards the blow off axis (indicated by the red dot on each plot) parallel to the the target surface, and the density gradient points toward the target surface ($\hat{-y}$). Thus, by $\nabla T_e \times \nabla n_e$ the detected azimuthal fields are consistent with being generated by the Biermann Battery effect.

\begin{figure}
    \centering
    \includegraphics[width = 0.48\textwidth]{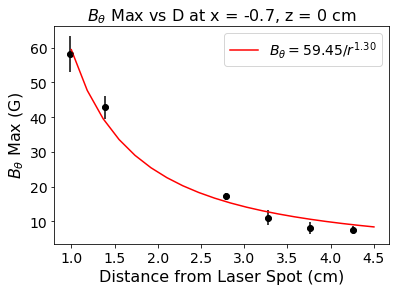}
    \caption{Plot of maximum azimuthal magnetic field vs. distance from the laser spot. The times at which each point occurs correspond to the times in Figure \ref{BmaxTime}. A $1/r^a$ curve (red line) agrees well with the data with a value of $a=1.3$.}
    \label{BvsD}
\end{figure}

The magnitude of the largest azimuthal magnetic field detected in each plane decreases with increasing distance from the target (Figure \ref{Planes}a). The maximum azimuthal magnetic field values for all planes versus the distance to the plane is shown in Figure \ref{BvsD}. A $1/r^a$ fit applied to the data shows good agreement. The inverse distance fit indicates that there is a $1/r^{1.3}$ spatial decay for the maximum magnetic field values as a function of distance from the target.

We calculate the current density $J_y$ normal to each $x$-$z$ plane of data using Ampere's law, $J\propto\nabla\times B$. The current densities are shown in Figure in \ref{Planes}b, corresponding to the same parameters as Figure \ref{Planes}a. Initially, the current density within each plane flows in the direction of plasma propagation (not shown). At times corresponding to the detection of Biermann fields, current begins to flow in both the positive (red) and negative (blue) directions along the blow-off axis. This current flowing toward the target surface begins near the origin of the plane and expands radially outward along with the Biermann fields. This current structure suggests that a current loop has formed, with the central current acting as a return current. Our measurement planes are too course to allow calculations of three-dimensional current structures, although these will be pursued in future experiments.

\begin{figure}
    \centering
    \includegraphics[width = 0.48\textwidth]{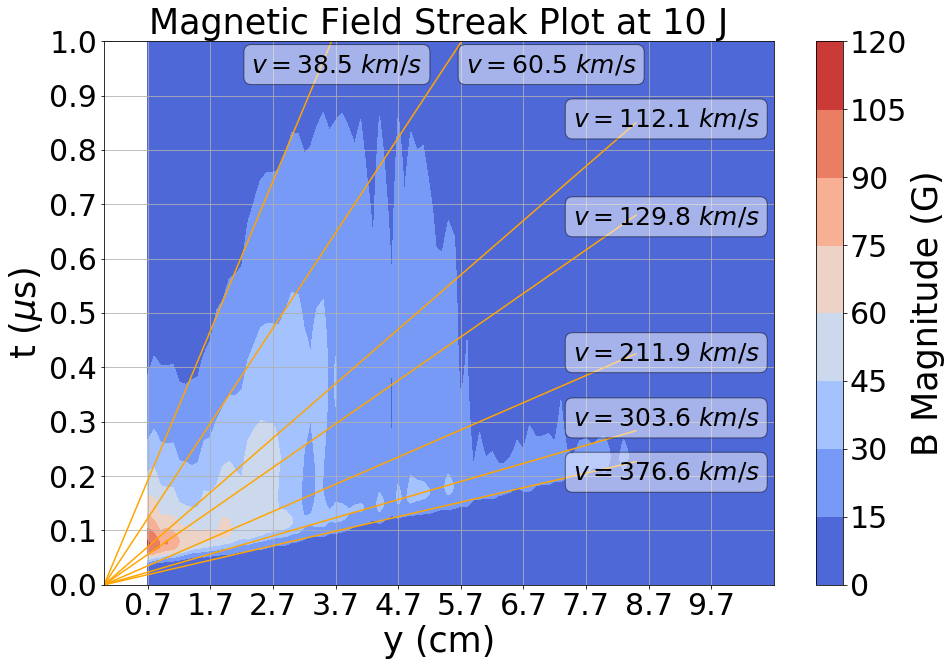}
    \caption{A streak plot of the total magnetic field (contour) from a $y$-lineout at $x = -0.7$ mm, $z = 0$ mm. Linear fits (orange lines) are applied to features of the magnetic field streak plot to determine the speed of different magnetic field features.}
    \label{Streak}
\end{figure}

\begin{figure}
    \centering
    \includegraphics[width = 0.48\textwidth]{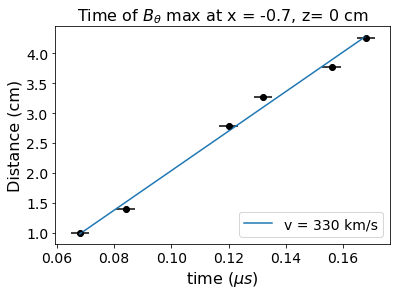}
    \caption{Plot of the maximum of the azimuthal magnetic field observed on the magnetic flux probe (black) at different planes as a function of time. A linear fit (blue line) to the data indicates a speed of 330 km s$^{-1}$.}
    \label{BmaxTime}
\end{figure}

Figure \ref{Streak} shows a streak plot of the magnetic field at $x=-0.7$ mm.  By applying linear fits to different features in the plot, the expansion speed of the magnetic fields was estimated to be between 300 - 370 km s$^{-1}$.  This is consistent with the expansion speed estimated from time-of-flight of the peak magnetic field at each measurement plane, which yielded a speed of approximately 330 km s$^{-1}$ as shown in Fig.~\ref{BmaxTime}.  
The expansion velocity of the laser plasma was calculated using the analytical model of Shaeffer et al.\cite{Schaeffer2016Characterization}, which yielded $v \approx 300 \pm 50$~km $s^{-1}$. This expansion velocity is consistent with the experimentally-determined speed of the magnetic fields, which in turn indicates that the magnetic fields propagate along with the plasma bubble via advection, as discussed below.

Electron temperature and density values were measured using an optical Thomson scattering diagnostic. A single data point at a spatial position of y = 1.5~cm from the heater beam spot along the blow-off axis shows $T_e = 10 \pm 2$~eV and $n_e = (5.55 \pm 1) \times 10^{16}\ cm^{-3}$.  Using the Thomson scattering and magnetic field data, we directly calculate the magnetic Reynolds number via Eq.~\ref{Rm} to be $R_m \approx 1.4\times 10^4$ at $y = 1.5$ cm. This indicates that, at this point in the system, advection with the plasma fluid flow dominates the propagation of magnetic fields. Additional Thomson scattering measurements are required to determine if advection is always the dominant process, if diffusion initially plays a role in the movement of the magnetic fields, and where and under what conditions a transition between diffusion and advection occurs. In future experiments, the optical Thomson scattering diagnostic will be extended into two dimensions to allow for measurements of electron temperature and density gradients at all spatial points within the system.

\section{\label{Sims} Numerical modeling with FLASH}

\begin{figure}
    \centering
    \includegraphics[width=0.48\textwidth]{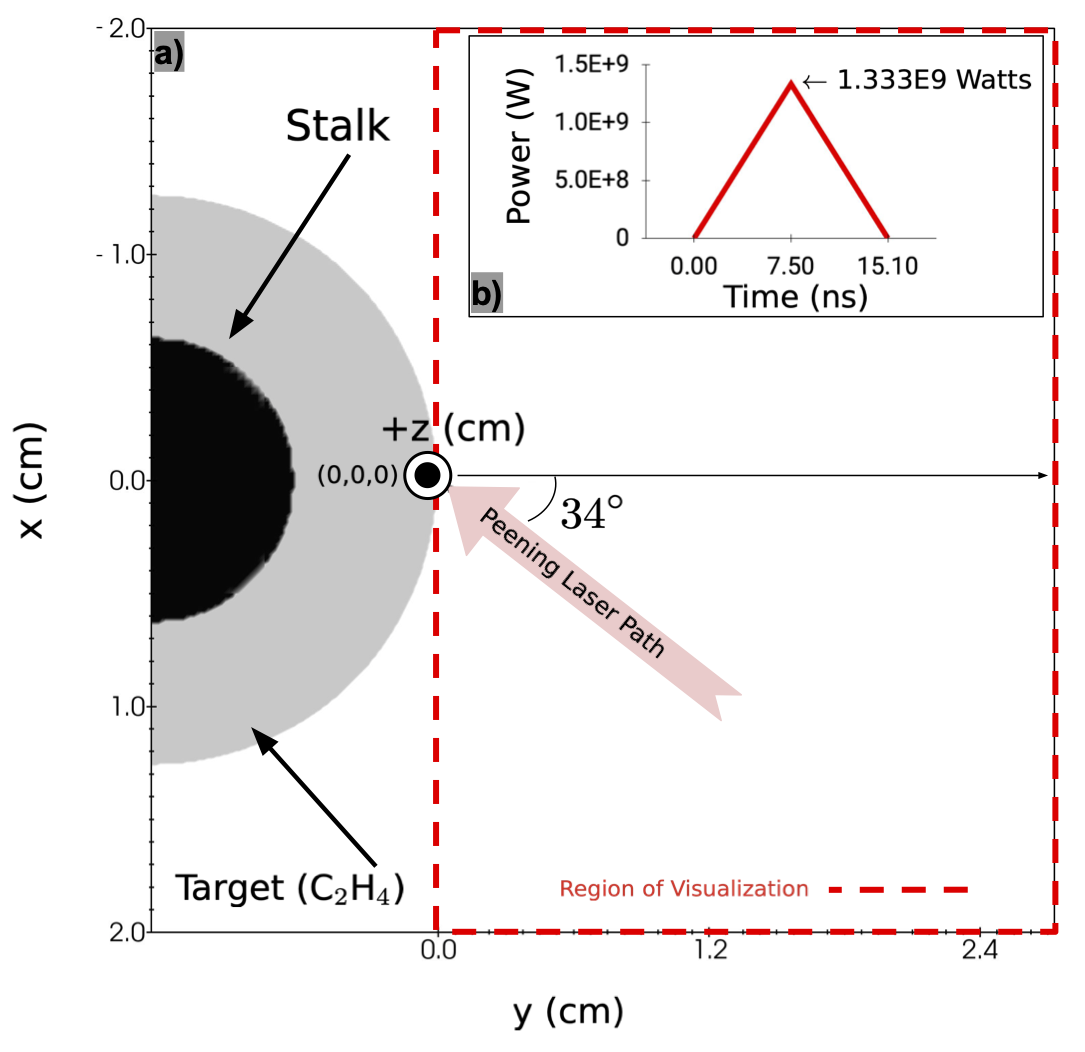}
    \caption{\textbf{a)} Visualization of the two-dimensional simulation domain for the $xy$-plane, i.e., $z=0$, at $t=0$ for the laser-facing side of the target. The black semi-circle region denotes the rod which supports the target material (grey). The Peening laser beam enters the simulation domain at a 34$^{\circ}$ angle from the $+\hat{y}$-direction for positive values of $\hat{x}$, reflecting the geometry of the experimental setup provided in Fig.~\ref{ExpSetUp}a. The region visualized in the provided simulation results (Figs.~\ref{Simulation_MaterialProperties} \& \ref{Simulation_MagneticFields}) is enclosed by the dashed red line. \textbf{b)} The power profile used to model the Peening laser heater beam in FLASH with a peak of $1.333\times10^9$ W at 7.5~ns, which allows 10~J of energy to be deposited to the target over 15~ns as in the experiment.}
    \label{Simulation_Initialization}
\end{figure}

\begin{figure}
    \centering
    \includegraphics[width=0.48\textwidth]{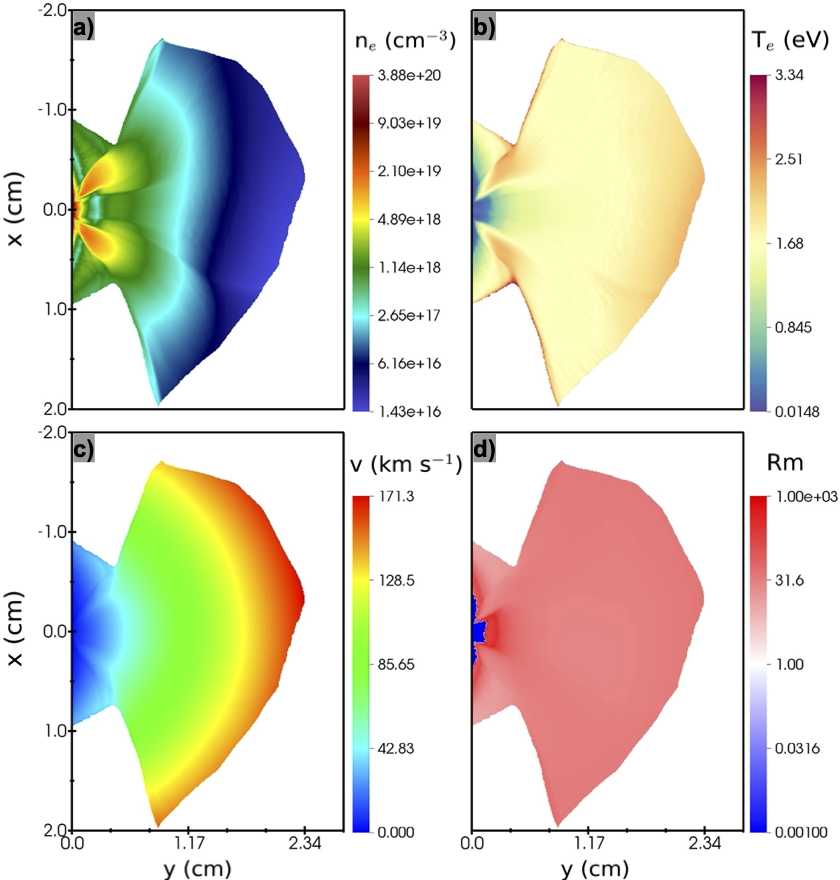}
    \caption{Visualization of the FBB 2D FLASH simulation for \textbf{a)} the electron number density $n_{e}$, \textbf{b)} the electron temperature $T_{e}$ \textbf{c)} the magnitude of the velocity, and \textbf{d)} the magnetic Reynolds number at 150 ns after the laser fires. We describe the threshold applied to these visualizations in the text.}
    \label{Simulation_MaterialProperties}
\end{figure}

\begin{figure*}
    \centering
    \includegraphics[width=\textwidth]{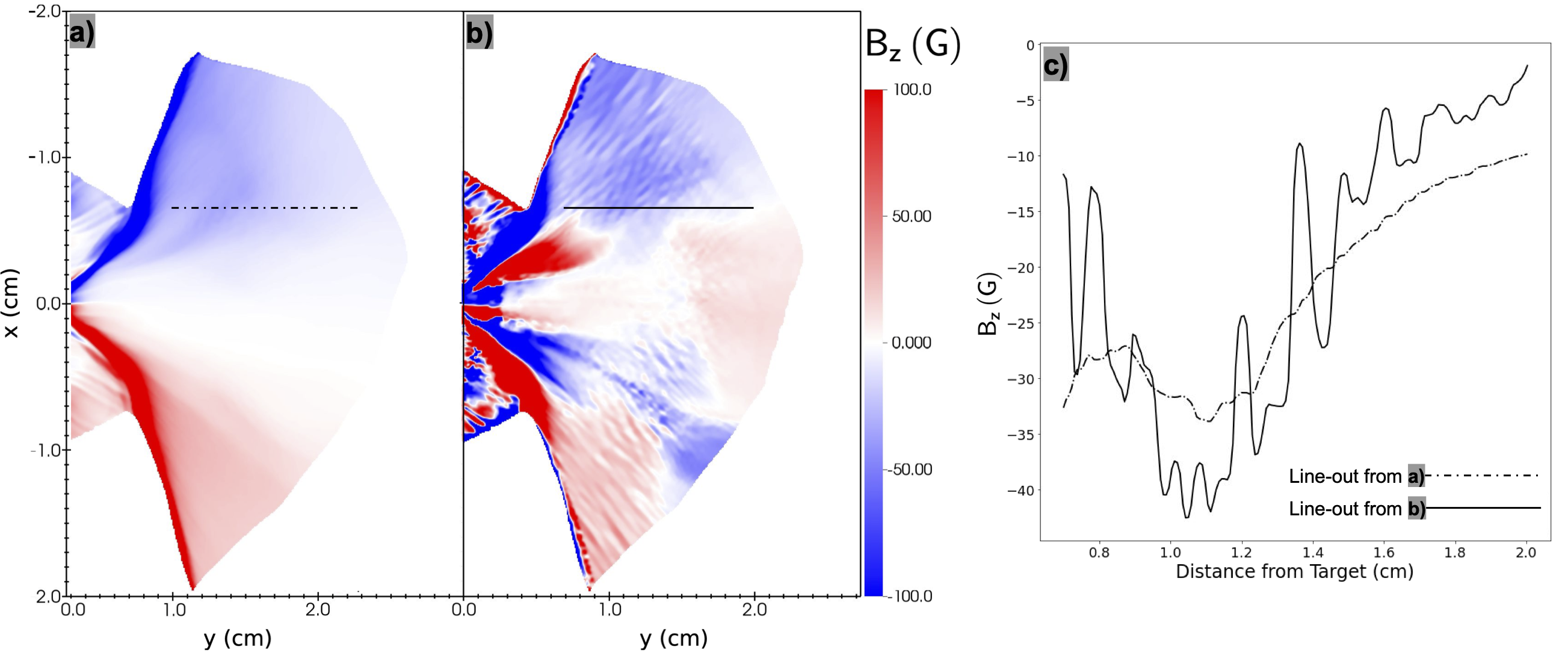}
    \caption{A visualization of the magnetic field values within the LPP region 150~ns after laser fire. \textbf{a)} results from a simulation where the Biermann battery source term was calculated only during the 15~ns duration of the laser (LOBB case), and \textbf{b)} results where the Biermann battery source term was calculated for the entire simulation duration of 400~ns (FBB case). Provided in \textbf{c)} are line-outs from a) LOBB and b) FBB simulations taken at $x=-0.7$~cm and $y=[0.7,2.0]$~cm away from the target.}
    \label{Simulation_MagneticFields}
\end{figure*}

Ancillary to future measurements that will explore the mechanisms behind the observed return current, and to further interrogate the magnetic field structure in the LPP, we have undertaken a series of simulations using the FLASH code\cite{Fryxell2000FLASH}. 
FLASH is a parallel, multi-physics, adaptive-mesh-refinement, finite-volume Eulerian hydrodynamics and MHD\cite{Lee2013} code, whose high energy density physics capabilities\cite{TzeferacosHEDP2015} have been validated through benchmarks and code-to-code comparisons~\cite{fatenejad2013collaborative, orban2013radiation}, as well as through direct application to laser-driven laboratory experiments\cite{meinecke2014turbulent,  meinecke2015developed, li2016scaled, tzeferacos2018laboratory, chen2020transport, bott2021time}.

The two-dimensional Cartesian simulation is initialized from a ``top-down''-perspective of the experimental configuration shown in Fig.~\ref{ExpSetUp}a. The simulation domain is illustrated in Fig.~\ref{Simulation_Initialization}a, modeling the $x$-$y$ plane (i.e., $z=0$) of the experiment. At room temperature and pressure, we initialize a cylindrical rod of C$_2$H$_4$ plastic. The equation of state and opacity material tables for C$_2$H$_4$ are computed using PrOpacEOS\footnote{For more information on PrOpacEOS, visit: \url{https://www.prism-cs.com/Software/Propaceos/overview.html}}. The rod initial mass density is $\rho = 1.047$ g/cc. We approximate the temporal profile of the laser pulse using a triangular profile, 15 ns long, with a peak power of approximately $1.33\times10^{9}$ W at 7.5 ns, as shown in Fig.~\ref{Simulation_Initialization}b. This drive profile emulates the 10 J Peening laser drive used in collecting the magnetic field data. The laser enters from the $+\hat{x}$-direction of the simulation domain at 34 degrees with respect to the $y$-axis, to accurately capture the incidence angle of the experimental drive.

In our finite-volume, single-fluid Eulerian simulations, the vacuum surrounding the rod must be modeled using a low density gas.  In order to image only the plasma expanding from the target rod, in Figs.~\ref{Simulation_MaterialProperties} \& \ref{Simulation_MagneticFields} we applied two threshold filters to visualize only the LPP properties against a white background. First, we exclude from the visualization cells containing a mass fraction less than 95\% of the rod material. Then, we exclude cells whose effective ionization $\bar{Z}$ is affected by heat flux from the compressed low-density gas material at neighboring cells. These thresholds exclude cells compromised by the presence of the low density gas from being folded into post-processing calculations. Similarly, magnetic field generation due to Biermann battery is only computed in smooth-flow regions of the LPP to ensure resolution-convergent magnetic field values\cite{fatenejad2013modeling}. 

We feature two simulation configurations that aim to determine (1) whether the magnetic fields measured experimentally are consistent with  Biermann battery generated magnetic fields, and (2), if so, to quantify the contribution of Biermann battery magnetic field generation in the expanding LPP versus the Biermann battery magnetic fields generated due to the laser-target interaction, which are subsequently advected by the expanding plasma. 

In the first simulation, we retain the Biermann battery source term in the induction equation operating throughout the entire simulation duration (i.e., ``Full Biermann battery'' or FBB). In the second simulation, we artificially switch off the Biermann battery source term as soon as the laser pulse ends (i.e., ``Laser Only Biermann battery'' or LOBB).
The plasma properties of the ``realistic'' FBB case are reported in Fig.~\ref{Simulation_MaterialProperties}, whereas the resulting magnetic field profiles for both FBB and LOBB simulations are shown in Fig.~\ref{Simulation_MagneticFields}. 
In Fig.~\ref{Simulation_MaterialProperties} we report the plasma properties of the LPP predicted by the FLASH FBB simulation. At the region of interest (i.e., the locus of the Thomson scattering measurements), we find electron number densities on the order of $6\times10^{16}$~cm$^{-3}$ (Fig.~\ref{Simulation_MaterialProperties}a) that match well with experimentally obtained values discussed in section IV. However, since the FLASH simulations have not yet been calibrated against the experimental results, we find that the simulation under-predicts the plasma electron temperatures and velocities (as shown in Fig.~\ref{Simulation_MaterialProperties}b,c), and is consistent with the experimental measurements only within a factor of unity. More specifically, the LPP expansion velocity in the FBB FLASH simulations is approximately 115-125~km s$^{-1}$ on average with a peak value of around 170~km s$^{-1}$, when the plasma velocity is estimated to be 330~km s$^{-1}$ in experiment. Consistent with the experimental results, we find that the LPP has magnetic Reynolds numbers $\mathrm{Rm} \gg 1$ (Fig.~\ref{Simulation_MaterialProperties}d), indicating that the magnetic field advection dominates resistive diffusion.

We note the emergence of two plasma lobes near the locus of the laser drive, which are prominently seen in the visualizations of the electron density and temperature (Fig.~\ref{Simulation_MaterialProperties}a,b). The two lobes surrounding the laser-target spot at the origin and the overall asymmetry of the LPP are caused by the asymmetry of the laser drive via the non-normal incidence angle. 
One may orient themselves conceptually by considering the bottom most density lobe ($+y$-direction) on the laser-facing side of Fig.~\ref{Simulation_MaterialProperties}a. This ejection has expanded more quickly in comparison to its data-collection side ($-y$-direction) sibling, as hot, dense material from the target has filled the comparably low density pseudo-vacuum bore from the simulated laser. These density and temperature gradients cause the generation of Biermann battery magnetic fields, and begin nanoseconds after the laser illuminates the target.
This is seen in both LOBB and FBB cases (panels a) and b) in Fig.~\ref{Simulation_MagneticFields}, respectively). The two lobes surrounding the laser-target interaction region have the strongest gradients, and thus the strongest magnetic fields values in the computational domain. 

Major points of comparison between panels a) and b) of Fig.~\ref{Simulation_MagneticFields} and the two resulting lineouts, taken at $x=-0.7$ cm to match the location of the experimental B-dot probe (Fig.~\ref{Simulation_MagneticFields}c), are the magnitude and structure (viz., spatial variability) of the Biermann battery magnetic fields. First, the values of the magnetic fields within the simulated LPP region are consistent and within range of the experimentally-measured values (Fig.~\ref{Planes}a), and are of the same direction. This observation further supports the conclusion that the experimentally-observed magnetic fields originate from the Biermann battery mechanism. Second, the magnetic fields featured in Fig.~\ref{Simulation_MagneticFields}a for the LOBB FLASH simulation are smooth when compared to those obtained in the FBB FLASH simulation (Fig.~\ref{Simulation_MagneticFields}b) and the experimentally-obtained magnetic field profiles (Fig.~\ref{Planes}a). By maintaining the Biermann battery source term active for the entirety of the simulation, the resulting magnetic fields manifest increased variability in both large (Fig.~\ref{Simulation_MagneticFields}b) and small spatial scales (Fig.~\ref{Simulation_MagneticFields}c). These large-scale spatial gradients of the Biermann battery magnetic fields can therefore naturally account for the current structures seen in Figure \ref{Planes}b, thus indicating that the Biermann battery mechanism in the experiment is active inside the LPP, even after the laser has fired. 
These small-scale structures in Fig.~\ref{Simulation_MagneticFields}b, seen also in the oscillations present in the solid line of Fig.~\ref{Simulation_MagneticFields}c are the result of spatial variations in the misalignment of electron temperature and density gradients, which result in continuous Biermann battery magnetic field generation, in contrast to  Fig.~\ref{Simulation_MagneticFields}a. Kinetic magnetic field generation effects such as the Weibel instability cannot be modeled in FLASH, which is an Eulerian finite-volume MHD code with isotropic pressure.

We speculate that the quantitative discrepancy between the simulation results and the experimental measurements of the electron temperature and velocity can be attributed to the following factors that degrade simulation fidelity: (1) The laser-power temporal profile in the simulation is roughly approximated by a triangular pulse, which alters the deposition rate and the flow dynamics; (2) the FLASH simulations need to be calibrated \emph{a posteriori} using experimental data to be able to reproduce the laser intensity of the experiment; (3) The spatial resolution of the 2D simulation ($80\,\mu$m) is not sufficient to resolve well the laser spot size ($\sim 200\,\mu$m). We are currently executing a simulation campaign with a series of two- and three-dimensional simulations at high resolution to calibrate the laser deposition and further validate the simulation results in a companion paper.

\section{Conclusions}
We have developed a high-repetition-rate experimental platform to examine the magnetic fields induced by laser-produced plasmas over large spatial regions. Data were taken in planes between 0.7 - 4.2 cm from the target surface, revealing azimuthal magnetic fields that reached a peak value of 60 G in the measurement plane closest to the target. The observed fields are azimuthally symmetric, consistent with a simple model of Biermann battery field generation in a cylindrically symmetric LPP. Based on the magnetic Reynolds number, $R_m$, directly calculated from data at $y = 1.5$ cm, transport of the self-generated fields away from the target surface is dominated by advection (rather than diffusion). However,  Biermann fields could continue to be generated where the fields are measured, as indicated by FLASH simulations of the experiment. 

The current density along the blow-off axis was calculated in each transverse plane via Ampere's law. The structure of the calculated currents indicates that there is a current loop that forms coincident with the Biermann fields. Further measurements are needed in order to calculate three dimensional current densities in the system.

Optical Thomson scattering was used to provide measurements of the plasma electron temperature and density at $y=1.5$~cm from the target surface along the blow-off axis. We are currently working to expand the scanning capabilities of the Thomson scattering diagnostic in order to obtain data in two dimensional planes, which will allow us to determine temperature and density gradients. Measurements of gradients will give us a direct comparison to MHD theory and will allow us to calculate $R_m$ for all spatial points in our system.

Future experiments will continue to probe the evolution of the Biermann battery fields over large volumes, allowing us to map 3-D regions of spontaneous field generation, as well as regions of advection and diffusion-dominated propagation within the system. Data collection will be expanded to planes further from the target surface and into the gaps between the existing data planes. In addition to increasing the resolution of data collection and expanding the volume over which we probe, we plan to investigate the effects of background gasses on spontaneous magnetic field generation. 

Further calibration and validation of our FLASH simulations are currently underway. This work will yield improved insights into (1) the coupling of the energy to the target, (2) the magnetic field morphology, (3) the observed potential return current and development of a potential current loop, and (4) set the foundation to study how the plasma properties produced in our experiments affect Biermann battery magnetic field generation. The simulation campaign will also provide a framework for platform design for future experiments.

\section*{Acknowledgements}
This work is supported by the Department of Energy (DOE) under award number DE-SC0019011, the National Nuclear Security Administration (NNSA) Center for Matter Under Extreme Conditions under Award Number DE-NA0003842 and the National Science Foundation Graduate Fellowship Research Program under award number DGE-1650604. The Flash Center for Computational Science acknowledges support by the U.S. DOE NNSA under Subcontracts No. 536203 and 630138 with Los Alamos National Laboratory, Subcontract B632670 with LLNL, and support from the Cooperative Agreement DE-NA0003856 to the Laboratory for Laser Energetics University of Rochester. We thank the University of Rochester's Center for Integrated Research Computing (CIRC). We also thank NIWC Pacific and Curtiss-Wright MIC for help with the slab laser system.


\begin{thebibliography}{00}
\bibitem{Kulsrud2008On} R.~M. Kulsrud and E.~G. Zweibel, \emph{Reports on Progress in Physics} \textbf{71}, 4 046901 (2008).

\bibitem{Rand1989The} R.~J. Rand and S.~R. Kulkarni, \emph{The Astrophysical Journal} \textbf{343}, 760 (1989).

\bibitem{Beck2007Magnetic} R. Beck, \emph{EAS Publications Series} \textbf{23}, p. 19--36 (2007).

\bibitem{Han2006Pulsar} J.~L. Han, R.~N. Manchester, A.~G. Lyne, G.~J. Qiao and W. van Straten, \emph{The Astrophysical Journal} \textbf{642}, 2 p. 868--881 (2006).

\bibitem{Zweibel1997Magnetic} E.~G. Zweibel and C. Heile, \emph{Nature} \textbf{385}, 6612 p. 131--136 (1997).

\bibitem{Heiles1998Zeeman} C. Heiles, \emph{Astrophysical Letters and Communications} \textbf{37}, p. 85--107 (1998).

\bibitem{Remington2000A} B.~A. Remington, R.~P. Drake, H. Takabe and D. Arnett, \emph{Physics of Plasmas} \textbf{7}, 5 p. 1641--1652 (2000).

\bibitem{Ryutov2001MHD} D.~D. Ryutov, B.~A. Remington, H.~F. Robey and R.~P. Drake, \emph{Physics of Plasmas} \textbf{8}, 5 p. 1804--1816 (2001).

\bibitem{Zweibel2013} E.~G. Zweibel, ``The Seeds of a Magnetic Universe,'' \emph{Physics} \textbf{6} (2013).

\bibitem{Naoz2013Generation} S. Naoz and R. Narayan, \emph{Physical Review Letters} \textbf{111}, 5 (2013).

\bibitem{Gregori2012Generation} G. Gregori, A. Ravasio, et al. 
\emph{Nature} \textbf{481}, 7382 (2012).

\bibitem{Biermann1949} L. Biermann, ``\"{U}ber den Ursprung der Magnetfelder auf Sternen und im interstellaren Raum,'' (in German) \emph{Zeitschrift Fur Naturforschung} (1949).

\bibitem{Stamper1971Spontaneous} J.~A. Stamper, K. Papadopoulos, R.~N. Sudan, S.~O. Dean, E.~A. McLean, and J.~A. Dawson, \emph{Physical Review Letters} \textbf{26}, 17 (1971).

\bibitem{Pisarczyk2015Space-time} T. Pisarczyk, S. Yu. Gus'kov, et al. 
\emph{Physics of Plasmas} \textbf{22}, 10 102706 (2015). 

\bibitem{Gopal2008Temporally} A. Gopal, M. Tatarakis, F.~N. Beg, E.~L. Clark, A.~E. Dangor, R.~G. Evans, P.~A. Norreys, M.~S. Wei, M. Zepf and K. Krushelnick, \emph{Physics of Plasmas} \textbf{15}, 12 122701 (2008).

\bibitem{Mckee1974Self-generated} L.~L. McKee and R.~S. Bird and F. Schwirzke, \emph{Physical Review A} \textbf{9}, 3 p. 1305--1311 (1974).

\bibitem{Stamper1991Review} J.~A. Stamper, \emph{Laser and Particle Beams} \textbf{9}, 4 p. 841--862 (1991).

\bibitem{Gao2015Precision} L. Gao, P.~M. Nilson, I.~V. Igumenshchev, M.~G. Haines, D.~H. Froula, R. Betti and D.~D. Meyerhofer, \emph{Physical Review Letters} \textbf{114}, 21 (2015).

\bibitem{Nilson2006Magnetic} P.~M. Nilson, L. Willingale, 
\emph{Physical Review Letters} \textbf{97}, 25 (2006).

\bibitem{Gregori2012Magnetic} G. Gregori and F. Miniati and B. Reville and R.~P. Drake, \emph{EAS Publications Series} \textbf{58}, p. 23--26 (2012).

\bibitem{Li2008Monoenergetic} C.~K. Li, F.~H. S\'{e}guin,  et al. 
\emph{Physical Review Letters} \textbf{100}, 22 (2008).

\bibitem{Walsh2017Self-Generated} C.~A. Walsh, J.~P. Chittenden, K. McGlinchey, N.~P.~L. Niasse and B.~D. Appelbe, \emph{Physical Review Letters} \textbf{118}, 15 (2017).

\bibitem{Campbell2020Magnetic} P.~T. Campbell, C.~A. Walsh, B.~K. Russell, J.~P. Chittenden, A. Crilly, G. Fiksel, P.~M. Nilson, A.~G.~R Thomas, K. Krushelnick, and L. Willingale, \emph{Physical Review Letters} \textbf{125}, 14 (2020).

\bibitem{Braginskii1965Transport} S.~I. Braginskii, ``Transport Processes in a Plasma,'' \emph{Reviews of Plasma Physics} \textbf{1}, 205 (1965).

\bibitem{Craxton1975Hot} R.~S. Craxton and M.~G. Haines, \emph{Physical Review Letters} \textbf{35}, 20 (1975).

\bibitem{Craxton1978JxB} R.~S. Craxton and M.~G. Haines, \emph{Plasma Physics} \textbf{20}, 6 p. 487--502 (1978).

\bibitem{Raven1978Megagauss} A. Raven, O. Willi, P.~T. Rumsby, \emph{Physical Review Letters} \textbf{41}, 8 (1978).

\bibitem{Haines1986Magnetic} M.~G. Haines, \emph{Canadian Journal of Physics} \textbf{64}, 8 p. 912--919 (1986).

\bibitem{McLean1984Observation} E.~A. McLean, J.~A. Stamper, C.~K. Manka, H.~R. Griem, D.~W. Droemer and B.~H. Ripin, \emph{Physics of Fluids} \textbf{27}, 5 1327 (1984).

\bibitem{Bird1973Pressure} R.~S. Bird, L.~L. McKee, F. Schwirzke and A.~W. Cooper, \emph{Physical Review A} \textbf{7}, 4 p. 1328--1331 (1973).

\bibitem{Dane1995Peening} C.~B. Dane, L.~E. Zapata, W.~A. Neuman, M.~A. Norton and L.~A. Hackel, \emph{IEEE Journal of Quantum Electronics} \textbf{31}, 1 p. 148--163 (1995).

\bibitem{Everson2009Design} E.~T. Everson, P. Pribyl, C.~G. Constantin, A. Zylstra, D.~B. Schaeffer, N.~L. Kugland and C. Niemann, \emph{Review of Scientific Instruments} \textbf{80}, 11 113505 (2009).

\bibitem{Schaeffer2018A} D.~B. Schaeffer, L.~R. Hofer, E.~N. Knall, P.~V. Heuer, C.~G. Constantin and C. Niemann, \emph{High Power Laser Science and Engineering} \textbf{6}, (2018).

\bibitem{Froula2012Plasma} D.~H. Froula, S.~H. Glenzer, N.~C. Luhmann, J. Sheffield and T.~J.~H. Donn\'{e}, \emph{Fusion Science and Technology} \textbf{61}, 1 p. 104--105 (2012).

\bibitem{Kaloyan2021Raster} M. Kaloyan, S. Ghazaryan, C.~G. Constantin, R.~S. Dorst, P.~V. Heuer, J.~J. Pilgram, D.~B. Schaeffer and C. Niemann, \emph{Review of Scientific Instruments} \textbf{92}, 9 093102 (2021)

\bibitem{Schaeffer2016Characterization} D.B Schaeffer, A.S. Bondarenko, E.T. Everson, S.E. Clark, C.G Constantin, and C. Niemann \emph{J. Appl. Phys.} \textbf{120} 043301 (2016)

\bibitem{Fryxell2000FLASH} B. Fryxell, K. Olson, P. Ricker, F.~X. Timmes, M. Zingale, D.~Q. Lamb, P. MacNeice, R. Rosner, J.~W. Truran and H. Tufo, \emph{The Astrophysical Journal Supplement Series} \textbf{131}, 1 p. 273--334 (2000).

\bibitem{Lee2013} D. Lee, \emph{Journal of Computational Physics} \textbf{243}, p. 269--292 (2013).

\bibitem{TzeferacosHEDP2015} P. Tzeferacos, M. Fatenejad, N. Flocke, C. Graziani, G. Gregori and D.~Q. Lamb, D. Lee, J. Meinecke, A.~M. Scopatz and K. Weide, \emph{High Energy Density Physics} \textbf{17}, p. 24-31 (2015).

\bibitem{fatenejad2013collaborative} M. Fatenejad, B. Fryxell, J. Wohlbier, E. Myra, D. Lamb, C. Fryer, and C. Graziani, \emph{High Energy Density Physics} \textbf{9}, 1 p. 63--66 (2013).

\bibitem{orban2013radiation} C. Orban, M. Fatenejad, S. Chawla, S.~C. Wilks and D.~Q. Lamb, ``A Radiation-Hydrodynamics Code Comparison for Laser-Produced Plasmas: FLASH versus HYDRA and the Results of Validation Experiments'' \emph{arXiv preprint}, arXiv:1306.1584 (2013). 

\bibitem{meinecke2014turbulent} J. Meinecke, H.~W. Doyle, et al. 
\emph{Nature Physics} \textbf{10}, p. 520--524 (2014).

\bibitem{meinecke2015developed} J. Meinecke, P. Tzeferacos, et al. 
\emph{Proceedings of the National Academy of Sciences} \textbf{112}, 27 p. 8211--8215 (2015).

\bibitem{li2016scaled} C.~K. Li, P. Tzeferacos, et al. 
\emph{Nature Communications} \textbf{7}, 13081 (2016). 

\bibitem{tzeferacos2018laboratory} P. Tzeferacos, A. Rigby, et al. 
\emph{Nature Communications} \textbf{9}, 591 (2018).

\bibitem{chen2020transport} L.~E. Chen,  A.~F.~A. Bott, P. Tzeferacos, A. Rigby, A. Bell, R. Bingham, C. Graziani, J. Katz, M. Koenig, C.~K. Li, \emph{The Astrophysical Journal} \textbf{892}, 2 (2020).
 
\bibitem{bott2021time} A.~F.~A. Bott, P. Tzeferacos, et al. 
\emph{Proceedings of the National Academy of Sciences} \textbf{118}, 11 (2021).

\bibitem{fatenejad2013modeling} M. Fatenejad, A.~R. Bell, et al. 
\emph{High Energy Density Physics} \textbf{9}, 1 p. 172--177 (2013).
\end{thebibliography}
\end{document}